\documentclass[fleqn,usenatbib]{mnras}

\usepackage{newtxtext,newtxmath}
\usepackage{graphicx}
\usepackage{dcolumn}
\usepackage{bm}
\usepackage{amsmath}
\usepackage{overpic}
\usepackage{booktabs}%
\usepackage{makecell}
\usepackage{enumerate}
\usepackage{nameref}
\usepackage{hyperref}
\usepackage{indentfirst}
\usepackage{float}
\usepackage{subcaption}
\usepackage{caption}
\usepackage{multirow}
\usepackage{caption}
\usepackage{threeparttable}
\usepackage{adjustbox}

\usepackage[T1]{fontenc}

\DeclareRobustCommand{\VAN}[3]{#2}
\let\VANthebibliography\thebibliography
\def\thebibliography{\DeclareRobustCommand{\VAN}[3]{##3}\VANthebibliography}

\usepackage{graphicx}	
\usepackage{amsmath}	

\title[GECAM discovery of the MXB/FRB 221014]{GECAM discovery of the second FRB-associated Magnetar X-ray Burst from SGR J1935+2154}

\author[C.W. Wang et al.]{
Chen-Wei Wang,$^{1,2}$ Shao-Lin Xiong,$^{1}$\thanks{E-mail: xiongsl@ihep.ac.cn}
Yue Wang,$^{1,2}$ Wen-Jun Tan,$^{1,2}$ Xiao-Bo Li,$^{1}$ Dong-Zi Li,$^{3}$ 
\newauthor
Yan-Qiu Zhang,$^{1,2}$ Shu-Xu Yi,$^{1}$ Ming-Yu Ge,$^{1}$ Sheng-Lun Xie,$^{1,4}$ Wang-Chen Xue,$^{1,2}$ Bing Li,$^{1}$  
Cheng-Kui Li,$^{1}$ 
 \newauthor
Zheng-Hua An,$^{1}$ Ce Cai,$^{5}$ Pei-Yi Feng,$^{1,2}$ Min Gao,$^{1}$ Ke Gong,$^{1}$ Dong-Ya Guo,$^{1}$ 
Hao-Xuan Guo,$^{1,6}$ Yue Huang,$^{1}$ 
 \newauthor
Jia-Cong Liu$^{1,2}$ Xin-Qiao Li,$^{1}$ Ya-Qing Liu,$^{1}$ 
Xiao-Jing Liu,$^{1}$ 
Xiang Ma,$^{1}$ Wen-Xi Peng,$^{1}$ Rui Qiao,$^{1}$ 
\newauthor
Yang-Zhao Ren,$^{1,7}$ 
Li-Ming Song,$^{1,2}$ 
Xi-Lei Sun,$^{1}$ Jin Wang,$^{1}$ Jin-Zhou Wang,$^{1}$ Ping Wang,$^{1}$ Xiang-Yang Wen,$^{1}$
 \newauthor
Shuo Xiao,$^{8,9}$ 
Sheng Yang,$^{1}$ Qi-Bin Yi,$^{10}$ Zheng-Hang Yu,$^{1,2}$ Da-Li Zhang,$^{1}$ Fan Zhang,$^{1}$
Wen-Long Zhang,$^{1,11}$ 
\newauthor
Jin-Peng Zhang,$^{1,2}$ Peng Zhang,$^{1,12}$ Shuang-Nan Zhang,$^{1}$ Zhen Zhang,$^{1}$ 
Xiao-Yun Zhao,$^{1}$ Yi Zhao,$^{13}$ 
 \newauthor
Chao Zheng,$^{1,14}$ Shi-Jie Zheng$^{1}$ 
\\
$^{1}$State Key Laboratory of Particle Astrophysics, Institute of High Energy Physics, Chinese Academy of Sciences, 19B Yuquan Road, Beijing 100049, China\\
$^{2}$University of Chinese Academy of Sciences, Chinese Academy of Sciences, Beijing 100049, China\\
$^{3}$Department of Astronomy, Tsinghua University, Beijing 100084, China\\
$^{4}$Institute of Astrophysics, Central China Normal University, Wuhan 430079, China\\
$^{5}$College of Physics and Hebei Key Laboratory of Photophysics Research and Application, Hebei Normal University, Shijiazhuang, Hebei 050024, China\\
$^{6}$Department of Nuclear Science and Technology, School of Energy and Power Engineering, Xi'an Jiaotong University, Xi'an 710049, China\\
$^{7}$School of Physical Science and Technology, Southwest Jiaotong University, Chengdu 611756, China\\
$^{8}$School of Physics and Electronic Science, Guizhou Normal University, Guiyang 550001, China\\
$^{9}$Guizhou Provincial Key Laboratory of Radio Astronomy and Data Processing, Guizhou Normal University, Guiyang 550001, China\\
$^{10}$School of Physics and Optoelectronics, Xiangtan University, Xiangtan 411105, China\\
$^{11}$School of Physics and Physical Engineering, Qufu Normal University, Qufu, Shandong 273165, China\\
$^{12}$College of Electronic and Information Engineering, Tongji University, Shanghai 201804, China\\
$^{13}$School of Computer and Information, Dezhou University, Dezhou 253023, Shandong, China\\
$^{14}$TIANFU Cosmic Ray Research Center, Chengdu, Sichuan, China
}

\date{Accepted XXX. Received YYY; in original form ZZZ}

\pubyear{\the\year{}}

\begin{document}
\label{firstpage}
\pagerange{\pageref{firstpage}--\pageref{lastpage}}
\maketitle

\begin{abstract}
Fast radio burst (FRB) is mysterious phenomenon with millisecond-duration radio pulses observed mostly from cosmological distance. 
The association between FRB 200428 and a magnetar X-ray burst (MXB) from SGR J1935+2154 has significantly advanced the understanding of FRB and magnetar bursts. 
However, it is uncertain whether this association between MXB and FRB (i.e. MXB/FRB 200428) is genuine or just coincidental only based on this single event. 
Here we report the discovery of a bright ($\rm\sim7.6\times10^{-7}\,erg \cdot cm^{-2}$ in 1-250 keV) magnetar X-ray burst detected by GECAM on October 14th, 2022 (dubbed as MXB 221014) from SGR J1935+2154, which is associated with a FRB detected by CHIME and GBT. 
We conducted a detailed temporal and spectral analysis of MXB 221014 with GECAM data and find that it is a bright and typical ($T_{90}\sim$250\,ms) X-ray burst from this magnetar. 
Interestingly, we find two narrow X-ray pulses in the MXB, one of which temporally aligns with the main pulse of the FRB 221014 $\sim5.70$ ms latter than the peak time of FRB 221014), resembling the feature found in MXB/FRB 200428. 
Furthermore, we did comprehensive comparison between MXB/FRB 221014 and MXB/FRB 200428, and find that while the two events share several common features, they also exhibit distinct differences, highlighting the variety of the MXB-FRB association morphology. 
This finding not only confirms the association between MXB and FRB but also provides new insights into the mechanism of and the relationship between FRB and MXB.
\end{abstract}




\section{Introduction}

Magnetar is a kind of neutron star with spin period ($P$) and fast spin-down rate ($\dot{P}$), usually manifesting as soft gamma-ray repeater (SGR) or anomalous X-ray pulsar (AXP) \footnote{The SGR/AXP classification is kind of obsolete. More and more observations indicate that SGR and AXP have the same properties. Such classification is more attributable to historical factors.} \citep[e.g.][]{Kouveliotou_SGR,McGill_Magnetar_Catalog,mangetars_ARAA,Negro_2024}. Magnetar has an extremely powerful magnetic field, about 10$^{14}$ to 10$^{15}$ G, or even higher, which is the primary energy source of energetic radiation and burst activities.

As one of the most prolific magnetars to date, SGR J1935+2154 was first discovered by Swift in 2014 \citep{2014GCN_Swift,2016MNRAS.457.3448I}, while there are some candidate bursts before 2014 that are found in Fermi/GBM data \citep{1935_periodicity,Xie2024ApJ}. Since the discovery, several burst episodes of SGR J1935+2154 have been observed \citep[e.g.][]{1935_periodicity}, including at least one burst forest \citep{GBM_1935_burst_forest,NICER_1935_burst_forest} as well as many outbursts \citep{Younes_1935_outburst,Lin_1935_outburst} in recent years. 
Some studies argued that the activities of SGR J1935+2154 exhibit periodic window behavior with a periodicity of $\sim238$\citep{231_day} days or $\sim134$\citep{1935_periodicity,GECAM_MXB_catalog} days, although the significance is marginal.

The rotation period ($P$) of SGR J1935+2154 is 3.25\,s while the $\dot{P}$ is 3.45$\times10^{-11}$\,s$\cdot$s$^{-1}$\citep{1935_Spin_Evolution}. The surface dipole magnetic field strength $B_s\sim2.2\times10^{14}$ G and the spin-down age is $\sim3.6$ kyr \citep{2016MNRAS.457.3448I}. This magnetar was found to be located in the supernova remnant G57.2+0.8 \citep{SNR_1935_GCN}, whose distance is measured to be 6–15 kpc \citep{1935_dis_1,1935_dis_2,1935_dis_3,1935_dis_4,1935_dis_5}. \cite{1935_dis_4} found that the age of the host SNR G57.2+0.8 is $\geq2\times10^4$ yr.

Remarkably, SGR J1935+2154 is the first magnetar confirmed to produce Fast Radio Burst (FRB). 
FRB is a kind of intense millisecond-duration burst event in the radio band first discovered in 2007 \citep{FRB_discovery}. The origin and mechanism of FRB have emerged as one of the most significant puzzles in modern astronomy. 
The extragalactic origin has already been suggested for the Lorimer Burst (i.e., the first detected FRB) due to the large dispersion measure (DM) \citep{FRB_discovery}. 
With the development of high-time-resolution and high sensitivity radio instrumentation, as well as the implementation of an increasing number of sky surveys in the radio band, a large sample of FRBs data has been accumulated \citep[e.g.][]{FRBCAT,CHIME_FRB_catalog}. 
It is future revealed that at least some FRBs are repeating \citep[e.g.][]{repeating_FRB,repeating_FRB_2,repeating_FRB_3,repeating_FRB_4}. 
With the observation of host galaxies of more and more FRBs, there is a growing consensus that most FRBs have an extragalactic origin \citep[e.g.][]{zhang_2020_FRB_model,xiao_2021_FRB_model,review_2,review_3,review_1,CHIME_FRB_catalog,CHIME_FRB_host_catalog}. 
However, neither the multi-wavelength counterpart of FRBs nor the galactic FRBs was detected before FRB 200428.

As the first galactic FRB, FRB 200428 is a milestone event, which is detected by Canadian Hydrogen Intensity Mapping Experiment (CHIME) \citep{CHIME_200428} and Survey for Transient Astronomical Radio Emission 2 radio telescope (STARE2) \citep{START2_200428}, although it is shown that, even much brighter than typical radio emission of pulsars, RRATs and magnetars, FRB 200428 is much less energetic than extragalactic FRBs by orders of magnitude \citep{CHIME_200428,START2_200428,review_2}. 
At the time of FRB 200428, 
several high-energy telescopes, including \textit{Insight}-HXMT \citep{HXMT_200428,HXMT_0428_re}, INTEGRAL \citep{INTEGRAL_200428}, AGILE \citep{AGILE_0428}, and Konus-Wind \citep{KW_0428}, detected a bright magnetar X-ray burst (MXB\footnote{It was called X-ray burst and XRB was used to be the acronym, however, this acronym has already been used to refer to X-ray binary. To resolve this confusion, we suggest to use MXB to specifically represent the x-ray burst from magnetar.}).

Although many theoretical models have been proposed, the physical mechanisms underlying FRB remain mysterious. All these models can be categorized into several frameworks, such as pulsar-like models, GRB-like models, close-in models, far-away models, etc \citep[see e.g.,][and references therein]{ platts_2019_FRB_model,xiao_2021_FRB_model,zhang_2020_FRB_model,lyu_2021_FRB_model}. 
And many models invoke neutron stars as the source of FRBs, considering the rapid variability, the repeating nature, and other observed characteristics \citep[e.g.][]{NS_model_1,NS_model_2,NS_model_3,NS_model_4,NS_model_5,NS_model_6,NS_model_7,NS_model_8,NS_model_9,NS_model_10,NS_model_11,NS_model_12,NS_model_13,NS_model_14,NS_model_18,NS_model_21,NS_model_22,NS_model_23}, especially after the discovery of FRB 200428 and the associated magnetar X-ray burst \citep[e.g.][]{NS_model_15,NS_model_16,NS_model_17,NS_model_19,NS_model_20,NS_model_24,NS_model_25,NS_model_27}.
FRB 200428 was the first observed FRB with a confirmed galactic origin and the first confirmed FRB with a neutron star (magnetar) origin, which indicates that at least some FRBs originate from magnetars. 

Importantly, apart from the general association between the magnetar X-ray burst and FRB 200428, Insight-HXMT team first reported that two narrow pulses (or say narrow peaks) of the magnetar X-ray burst are well aligned with two radio pulses of FRB 200428 and suggested that the narrow pulses are very likely the high-energy counterpart of the FRB \citep{HXMT_0428_GCN,HXMT_200428}. Such alignment of x-ray and radio pulses is also seen in the INTEGRAL observation \citep{INTEGRAL_200428}. 
Dedicated analysis of Insight-HXMT data showed the lightcurve of MXB 200428 actually consists of several broad bump-like components and two narrow pulses, and the peak times of the two narrow X-ray pulses were slightly later than the peak times of two pulses of FRB 200428, respectively \citep{HXMT_0428_re}. 
Interestingly, there is a rather significant $\sim$40 Hz quasi-periodic oscillation (QPO) feature found in the light curve of MXB 200428 \citep{200428_QPO}, which is very rare in other MXBs from this magnetar. Also, the QPO frequency is well consistent with the waiting time between these two narrow pulses associated with FRB pulses.
These results demonstrate that not only the magnetar X-ray burst but also the narrow pulse in the magnetar X-ray burst could work as probes for studying the behavior and nature of the magnetar burst activity, as well as FRB.

Despite the great importance of the association between MXB and FRB, there is only one confirmed case (i.e., MXB/FRB 200428) detected before. 
However, it remains difficult to confirm which features are universal features of magnetar X-ray bursts associated with radio bursts and which are exclusive features of MXB 200428.
Here we report the discovery of the second MXB-FRB association from the prolific source SGR J1935+2154 on Oct. 14th, 2022. For convenience, we named this association event as MXB/FRB 221014. This MXB was detected and first reported by Gravitational wave high-energy Electromagnetic Counterpart All-sky Monitor (GECAM) \citep{GECAM_221014_ATEL}, and later confirmed by Konus-Wind, although its data resolution is limited as it was in the waiting mode \citep{KW_221014_ATEL}. 

This paper is structured as follows. Detailed observation of GECAM-B and GECAM-C is described in Section 2. 
In Section 3, we present the data reduction and analysis results of GECAM-B and GECAM-C. 
The summary and discussion are given in Section 4.

\section{Instrument and Observation}

GECAM is a constellation with four X-ray and gamma-ray wide FoV monitors, including GECAM-A and GECAM-B \citep{GEC_INS_Li2022}, GECAM-C (also called SATech-01/HEBS, \cite{HEBS_INS_Zhang2023}), and GECAM-D (also called DRO/GTM, \cite{GTM_INS_wang2024}). 
The primary science objectives of GECAM include GRBs \citep{09A_line,09A_afterglow,25A,07A_minijet,GECAM_overview}, SGRs \citep{1935_periodicity,GECAM_MXB_catalog,GECAM_overview}, high energy counterparts of Gravitational Wave Events and Fast Radio Bursts \citep{GECAM_overview,type_IL,GECAM_221014_ATEL} and high-energy atmospheric phenomena \citep{atmo_3,atmo_2,atmo_1}. 

On October 14th, 2022, while GECAM-A was mostly offline due to power supply issue, GECAM-B and GECAM-C were operating normally. There are 25 gamma-ray detectors (GRDs) in each satellite of GECAM-A/B and 12 GRDs in GECAM-C with the similar design. Each GRD has two readout channels, including the high gain (HG) channel and the low gain (LG) channel \citep{GEC_INS_An2022}, with only exception of two GRDs of GECAM-C. The HG channel is used for the analysis of hard X-ray photons, which is about 40 to 300 keV for GECAM-B and 6 to 300 keV for GECAM-C \citep{B_calibration,C_calibration,cross_B,cross_C}, which covered almost all signals from the X-ray burst of SGR J1935+2154. The LG channel is designed to detect soft $\gamma$-ray photons above 300 keV. Both GECAM-B and GECAM-C have a high temporal resolution of 0.1 $\mu$s \citep{B_time,C_time}.

In October 2022, SGR J1935+2154 became active again and produced a series of bursts. And the rate of X-ray bursts reached a peak on October 14th. At 2022-10-14T19:21:39.100 (denoted as $T_0$, see detail in Section\,\ref{sec:result}), both GECAM-B and GECAM-C are triggered in-flight \citep{GECAM_trigger,GECAM_alert} and on-ground \citep{GECAM_search_2025,ETJASMIN_II} by a bright X-ray burst (denoted as MXB 221014) with a duration of about 250\,ms \citep{GECAM_221014_ATEL,GECAM_MXB_catalog}. 
Both the in-flight and on-ground localization \citep{GECAM_localization,GECAM_MXB_catalog,ETJASMIN_I,ETJASMIN_II} are well consistent with SGR J1935+2154 within the error.

At 2022-10-14T19:21:39.130 UTC (topocentric time at 400 MHz), CHIME detected a bright radio burst at 600 MHz with a fluence as high as 9.7$\pm$6.7 kJy ms. With the high brightness, this FRB-like event is named as FRB 20221014A (FRB 221014 hereafter \footnote{\cite{CHIME_221014_arxiv} named the FRB associated with MXB from SGR J1935+21541 on April 28th, 2020 as FRB 20200428D and the FRB-like event on October 14th, 2022 as FRB 20221014A. To simplify and maintain consistency with other work to avoid confusion, we abbreviate them as FRB 200428 and FRB 221014 respectively.}) by the CHIME team \citep{CHIME_221014_ATEL,CHIME_221014_arxiv}. 
It should be noted that although this radio burst is much more energetic than typical radio pulse emission from magnetars and typical pulsars, this event is much less energetic than FRB 200428, and significantly less energetic than other extragalactic FRBs. 
However we also caution that FRB 221014 occurred at the edge of the CHIME field of view, which may lead to an underestimation of its fluence \citep{CHIME_221014_arxiv}. Considering the time delay caused by the DM, this radio burst is temporally associated with the MXB 221014. 

Interestingly, 
Green Bank Telescope (GBT) also detected a series of bright radio bursts (at least 5 significant bursts within a time span of 1.5 seconds) at 5 GHz \citep{GBT_221014_ATEL}, including two brightest bursts occurred at 2022-10-14T19:21:39.1 (topocentric, infinite frequency arrival time), whose high brightness even resulted in severe saturation of the receiver of GBT. 
As the detailed observation results of GBT are not available yet, we focus on the CHIME results in this work.

NICER also conducted a series of observations of SGR J935+2154 from October 13th to 26th, 2022. Two glitches were detected in the persistent X-ray emission, and the phase of FRB 221014 roughly aligns with the valley of soft X-ray pulse phase \citep{1014_phase}.

\section{Results and Discussions} \label{sec:result}

Since there is no significant signal in the LG channel of GECAM for MXB 221014, only HG channel GECAM data are used for the analysis. 
Considering the impact of incident angle on measurement and the signal-to-noise ratio, we selected GRD 05, 06, 12, 13, 14, 15, 21, 22, 23, 24 of GECAM-B and GRD 05, 07, 11, 12 of GECAM-C for temporal analysis, while GRD 13, 14, 15, 22, 23 of GECAM-B and GRD 07, 11, 12 of GECAM-C for spectral analysis. All the GRDs used for spectral analysis have an incident angle smaller than 60$^\circ$.

The Bayesian Information Criterion (BIC) is used for both model comparison of spectral analysis and temporal analysis, which is defined as BIC=-2ln$L$+$k$ln$N$, where $L$ represents the maximum likelihood value, $k$ denotes the number of free parameters in the model, and $N$ signifies the number of data points. The model with the smallest BIC value is preferred.

\subsection{Temporal analysis}
As the travel time of light is not negligible for investigating the temporal properties of MXB 221014 and FRB 221014, all lightcurves, including hard X-ray from GECAM-B and GECAM-C, radio from CHIME, are converted to Barycentric Dynamical Time (TDB), as described in Table\,\ref{tab:barycentric_correction}. 
The lightcurves of MXB 221014 and FRB 221014 are depicted in Fig.\ref{fig:lc_mxb_frb}\,(a), with $T_0$ of 2022-10-14T19:21:39.100 (trigger time of GECAM-B in UTC, corresponding to MJD 59866.8081787363 in TDB). 

The pulse phase profile of persistent soft X-ray emission are also obtained from NICER observations (OBSID 5020560106-5020560109 and OBSID 5576010101-557601011), with the same reduction analysis with \cite{1935_Spin_Evolution}. 
It can be seen in Fig.\ref{fig:lc_mxb_frb}\,(d) that MXB 221014 occurs during the bridge emission in the soft X-ray pulse phase profile, as it roughly aligns with the valley, which is consistent with the results of MXB/FRB 220428. As the opposite phase between soft x-ray and hard x-ray, this also suggests that MXB 221014 generally aligns with the peak region of the hard X-ray pulse phase \citep{1014_phase}.

One may find that the apparent duration of MXB 221014 detected by GECAM is much shorter than that of MXB 200428 detected by \textit{Insight}-HXMT (especially compared with light curves of the Low Energy X-ray telescope and the Medium Energy X-ray telescope) \citep{HXMT_0428_GCN,HXMT_200428}, but this difference is largely attributed to the different energy range of these two instruments. Comparing their light curves in the same energy band, one can see that the intrinsic durations of these two events are very similar. 

One of the most important features of the MXB/FRB 200428 is the temporal alignment of their pulses in hard X-ray and radio bands \citep{HXMT_0428_GCN,HXMT_200428,INTEGRAL_200428}.
Therefore, whether this behavior is commonly presented in MXB-FRB associated events is an important question. 
As depicted in Fig.\ref{fig:lc_mxb_frb}\,(a), although MXB 221014 and FRB 221014 are temporally associated with each other, the peak time of FRB 221014 (from CHIME observation) clearly does not coincide with the peak region but the tail of MXB 221014, which significantly differs from the case of MXB/FRB 200428.  
Following the same analysis process of MXB 200428 \citep{HXMT_0428_re}, we noticed that, as shown in Fig.\ref{fig:lc_mxb_frb}(c) and Table.\ref{tab:pulse_fitting}, the lightcurve of MXB 221014, which is obtained by stacking data from GECAM-B and GECAM-C, can be fitted by two narrow Gaussian functions (denoted as P2 and P3) and two broad Gaussian functions (denoted as P1 and P4) with $\chi^2/dof\sim2138/1988$. 
And the X-ray narrow-pulse P2, whose profile is evident in 6-30 keV and more significant in 50-300 keV (Fig.\ref{fig:lc_mxb_frb}(a)), aligns well in time with the the radio pulse of FRB 221014 detected by CHIME, as depicted in Fig.\ref{fig:lc_mxb_frb}(a) and Fig.\ref{fig:lc_mxb_frb}(c). 
The time delay of the peak of P2 (aligns with the bright radio pulse) and P3 (aligns with the possible weak radio pulse) is 5.70$^{+1.32}_{-1.49}$\,ms and 43.80$^{+1.68}_{-1.59}$\,ms compared to the peak time of the bright radio pulse of FRB 221014 detected by CHIME, respectively. This indicates that the genuine X-ray pulse associated with this FRB main pulse should be the narrow Gaussian profile X-ray pulses (P2) rather than the broader main pulse (P1) or narrow pulse P3 of MXB 221014.

Moreover, we noticed that there is some marginal excess in the CHIME data (Fig.4 in \cite{CHIME_221014_arxiv}) during the time interval of narrow x-ray pulse P3 by visual checking, suggesting there is probably a weaker radio pulse following the main brighter radio pulse of this FRB.
Although the significance of the latter radio pulse in CHIME data is very low, the facts that GBT detected multiple pulses from this FRB and that the temporal coincidence between the possible weak radio pulse and the narrow x-ray pulse P3 somewhat increase the possibility.   
If so, one will find that these two pulses of FRB 221014 (one bright and one weak) detected by CHIME are well aligned with the two narrow pulses (P2 and P3) in MXB 221014 respectively. The detailed observation results of GBT will help to determine whether the possible weak radio pulse coincident with P3 is real. 

In any case, we can confirm the presence of at least one X-ray pulse (i.e. P2) in MXB 221014 that aligns with the radio pulse in FRB 221014. These temporal alignment and delay features of x-ray narrow pulse and radio pulse in MXB/FRB 221014 exactly resemble the case of MXB/FRB 200428 (see Fig.\ref{fig:lc_mxb_frb}(b) and Figure 1 in \cite{HXMT_0428_re}). However, the difference is that all the narrow x-ray pulses and radio pulses occur during the weak tail region of MXB 221014, while they are around the bright peak region of the MXB 200428.

Moreover, a rather significant Quasi-periodic Oscillation (QPO) with a frequency of $\sim$40\,Hz is found in the MXB 200428 \citep{200428_QPO}. And the mini-pulses of MXB 200428 that coincide with the radio pulses contribute to the peaks of the oscillations. 
Considering that the time interval of the two narrow pulses in MXB 221014 is comparable to that of the narrow pulses in MXB 200428 \citep{HXMT_0428_re}, We also conducted a search for the potential QPO signal in MXB 221014 to investigate whether significant QPO is a common feature of the FRB-associated MXBs, based on the Leahy Power Spectrum Density \citep{leahy,stringray_1,stringray_2} and Morlet Wavelet \citep{Wavelet,PyWavelets} in four energy channels (6-30\,keV, 30-50\,keV, 50-100\,keV and 100-300\,keV). However, no significant (i.e., over 3$\sigma$) QPO signal is observed in MXB 221014. 
Although limited by the sensitivity of the detector, the observation results of MXB 221014 cannot rule out the possibility of the existence of very weak QPO signals or QPO signals in the soft X-ray band. Nevertheless, this strongly suggests that the significant QPO in the hard X-ray band is not a necessary feature of MXBs associated with FRBs, which provides constrains on the numerous magnetar origin models of FRBs.

We also note that, both MXB/FRB 200428 and MXB/FRB 221014 are multi-pulse events, not only the MXB contains multiple narrow pulses, but also the FRB is composed of multiple radio pulses. Interestingly, in the x-ray band, the narrow pulses ride on the broad pulses which contribute the main emission (fluence and energy) of the MXB, however, there seems no broad pulses underlying the apparent pulses of these FRBs in the radio band. Such features may have important implications of the physics and mechanism of MXB and FRB.

\subsection{Spectral analysis}


Past studies have shown that the spectra of MXBs are varied. Some MXB are dominated by non-thermal emission while some are dominated by thermal emission \citep{HXMT_1935_catalog_2,GBM_catalog_2021_Jan,GBM_catalog_2022_Jan,GBM_catalog_2022_Oct}. 
Thus, we used a combination of spectral models, including CPL, bbodyrad (BB), PL, CPL+BB, PL+BB, and BB+BB, to fit the time-integrated spectrum of MXB 221014 from $T_0$-0.11\,s to $T_0$+0.12\,s with PGstat by \textit{XSPEC}. The fitting results are listed in Table\,\ref{tab:spec_fitting}. Note that the parameters of CPL+BB are not well constrained. The BIC indicates that the CPL model is the most favored with reasonable spectral parameters but a relatively high fluence, with an $E_{iso}$ of 7.35$^{+0.83}_{-0.65}\times10^{39}$\,erg by assuming the magnetar distance of 9 kpc. The fitting results are shown in Fig.\ref{fig:spec_fitting}\,(a). 

MXB is generally believed to be produced from a trapped fireball of photon-pair plasma in the magnetosphere, which originates from the sudden release of magnetic energy or crustal stress triggered by the magnetic reconnection or crust crack. 
The fireball will subsequently expand from lower altitudes to higher altitudes in the closed magnetospheric zone and cool.
With a high Thomson optical depth, the emission from the fireball will be Comptonized and manifest as a CPL spectrum. Thermalization may occur if the opacity is high enough, producing a BB component. Additionally, the magnetar surface can also be heated, leading to thermal emission. All these may contribute to the detectable thermal component in the spectrum of MXB \citep[see e.g.,][and references therein]{Gill_2010_MXB,Lyutikov_2003_MXB,Younes_2014_MXB,Thompson_1995}. 

A possible origin of the CPL spectrum is the mimic of unsaturated Comptonization. 
Photons are repeatedly scattered until their energy reaches the plasma temperature $kT_e$, resulting in a quasi-cutoff in the spectrum. 
The photo index below the cutoff can be obtained as 
$-1/2+\sqrt{(9/4+4/y_{\rm B})}$, with $y_{\rm B}=4kT_e/(m_ec^2){\rm max}\{\tau_{\rm B},\tau^2_{\rm B}\}$ for a non-relativistic situation, where $\tau_{\rm B}$ is the effective scattering optical depth \citep{Radiative_1986_Rybicki,Lin_0501,Younes_2014_MXB}. 
The photo index $\sim1$ only when $y_{\rm B}\gg1$, suggests a high opacity. 
In this case, the spectrum is likely to thermalize and produce a blackbody component. 
Since the best model for the time-integrated spectrum of MXB 221014 is a non-thermal CPL model, we further investigated the upper limit of the flux of the thermal component in MXB 221014, which can provide some constraints on the fireball. Otherwise, if the fireball is too large or the temperature is high enough, significant thermal components should be observed in the spectrum, and the CPL model would no longer be the best fit. 

Therefore we compared the BIC of the single CPL model and the CPL+BB model with different temperatures and fluxes. The result is shown in Fig.\ref{fig:spec_fitting}\,(b). 
The white line represents the contour where $\Delta$BIC equals 10, indicating that parameter combinations within the red region above this line are strongly disfavored. 
The parameter combinations that fall within the blue region cannot be rejected by the joint spectral fitting results of GECAM-B/C. Thus, this provided a series of upper limits of both the size and temperature of the fireball, as shown in Fig.\ref{fig:spec_fitting}\,(b) (e.g. the R$^2_{\rm fireball}$ needs to be smaller than 1.20 km$^2$ if the temperature is 10 keV when taking $\Delta\rm BIC=2$ as a criterion). 

\cite{NS_model_27} proposed a GRB-like model and predicted that the X-ray bursts associated with FRBs should include a non-thermal synchrotron radiation component as well as a thermal photospheric component, and this model can well describe MXB 200428. Considering that both MXB 221014 and MXB 200428 have similar fluence and lack significant blackbody components, it may be possible to describe them using the same model, though further detailed investigation is needed. 

On the other hand, we also note that a photo index of $\sim1$ for CPL means that the shape of the non-thermal CPL spectrum is nearly the same as that of the optically thin thermal bremsstrahlung (OTTB) spectrum in the energy range of GECAM, making them indistinguishable. 
Hence we can not completely rule out a thermal origin for the emission. 

As shown in Fig\ref{fig:spec_pars}\,(a), we compared the spectral properties of MXB 221014 and MXB 200428 to other MXBs from SGR J1935+2154 \citep{HXMT_1935_catalog_2,GBM_catalog_2021_Jan,GBM_catalog_2022_Jan,GBM_catalog_2022_Oct}, and find that these two FRB-associated MXBs show similarity: their spectral properties (Epeak, alpha and fluence of the CPL model) are not outliers in the sample of X-ray bursts, but they all tend to have higher fluence in the sample.

In most FRB models invoking magnetar, the FRBs are believed to be generated by coherent radiation, and could only be generated in the gap region in the poles of the magnetic field of magnetar, where the relativistic fireball is formed that can produce X-ray bursts. 
Thus, the coincidence of pulses in a non-thermal or comptonized hard X-ray burst and radio burst is expected, just like MXB 200428 and FRB 200428. If the pulses of MXB not align with those in FRB, the X-ray bursts are inclined to not come from the poles, and a significant thermal component should be observed. 
Given that MXB/FRB 221014 shows good temporal alignment and MXB 221014 has non-thermal spectrum,
the MXB/FRB 221014 should share a similar physical mechanism as MXB/FRB 200428.

Both the MXB/FRB 200428 and MXB/FRB 221014 are coincident with the peak of the phase profile of the magnetar persistent emission in hard X-ray band. 
The persistent emission of SGR J1935+2154 in the soft X-ray band is composed of a thermal component dominated in soft X-ray and a non-thermal component extended to hard X-ray. \cite{1014_phase} suggests that the persistent non-thermal hard X-ray emission originates from the quasi-equatorial region. 
As both of MXB 221014 and MXB 200428 have non-thermal spectra, in this scenario, the X-ray bursts may also come from a quasi-equatorial region, otherwise the X-ray burst may not coincide with the peak of the hard X-ray pulse profile. 
But the FRBs from the quasi-equatorial region may be difficult in pulsar-like models or close-in models. 

On the other hand, it is possible that hard X-ray persistent emissions, as well as X-ray bursts and FRBs, all originate from the polar region of magnetar. In these models, however, one must address why the peak region of the main and broad pulse of MXB 221014 is not associated the FRB (without consideration of the GBT-detected FRB radio pulses whose time is not published yet), which is opposite to the situation of MXB/FRB 200428. Specifically, in the magnetar-asteroid interaction model \citep{NS_model_16}, it is proposed that the two pulses in MXB/FRB 200428 are generated by two major fragments of an asteroid distorted tidally. If MXB/FRB 221014 is produced by the same process, it is necessary to consider how the asteroid fragments can avoid generating radio emission when producing the bright pulses of X-ray burst. 
One possible explanation is that the radio emission of FRB is highly beamed while the x-ray emission of MXB is quasi-isotropic. 
Alternatively, the radio emission of FRB and X-ray emission of MXB may originate from different locations, such as X-rays in the polar regions and radio emissions at the quasi-equatorial region \citep{NS_model_15}.

\subsection{Evolution of burst rate}

Based on the samples collected from GECAM \citep{GECAM_MXB_catalog}, \textit{Fermi}/GBM \citep{GBM_burst_sample} and \textit{Insight}-HXMT \citep{HXMT_1935_catalog_1}, the X-ray bursts of SGR J1935+2154 from April 2020 to April 2023 are presented in Fig.\ref{fig:Fig_burst_rate}. An interesting feature is that both MXB/FRB 200418 and MXB/FRB 221014 occurred during the decaying phase of the active period in 2020 and 2022, respectively, where the burst rate gradually decreases. Although this result cannot be completely ruled out as an observational effect, as many radio telescopes did not continuously monitor SGR J1935+2154 during the early active phase with no interruption in observational coverage, and we cannot exclude the possibility of undetected radio bursts during the intense burst phase. 
Such observation is still not as expected by some magnetar-asteroid interaction model \citep{NS_model_16}, in which scenario an X-ray burst forest led by the instability of the crust should follow the FRB-associated X-ray burst. But the rate of X-ray burst rapidly decreased after the FRB-associated X-ray burst, especially for MXB/FRB 221014, a sudden decrease in burst rate after this event can be seen in Fig.\ref{fig:Fig_burst_rate}\,(b).

\section{Summary and Conclusion}
The magnetar X-ray burst MXB 200428 is the only multi-wavelength counterpart of FRB discovered before. 
This one-of-a-kind sample with special properties leads to a trend of trying to build general models to explain all the unique and outlying features detected in MXB 200428, which are considered to be universal features of magnetar X-ray bursts associated with radio bursts. 

\begin{figure*}
\centering
\begin{tabular}{cc}
        \adjustbox{valign=c}{\begin{overpic}[width=0.3\textwidth]{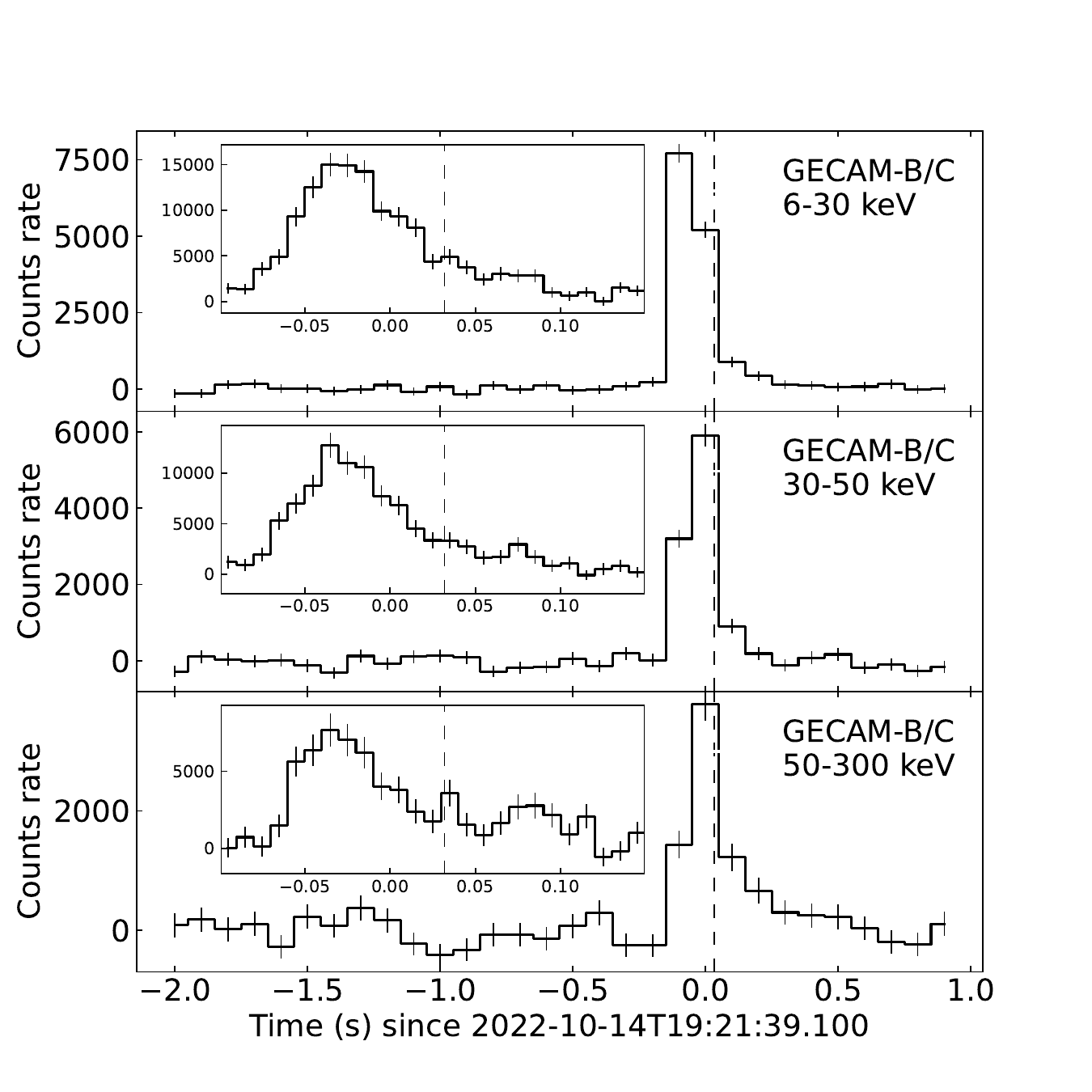}
            \put(-3, 90){\bf a}
        \end{overpic}}
        &
        \adjustbox{valign=c}{\begin{overpic}[width=0.2\textwidth]{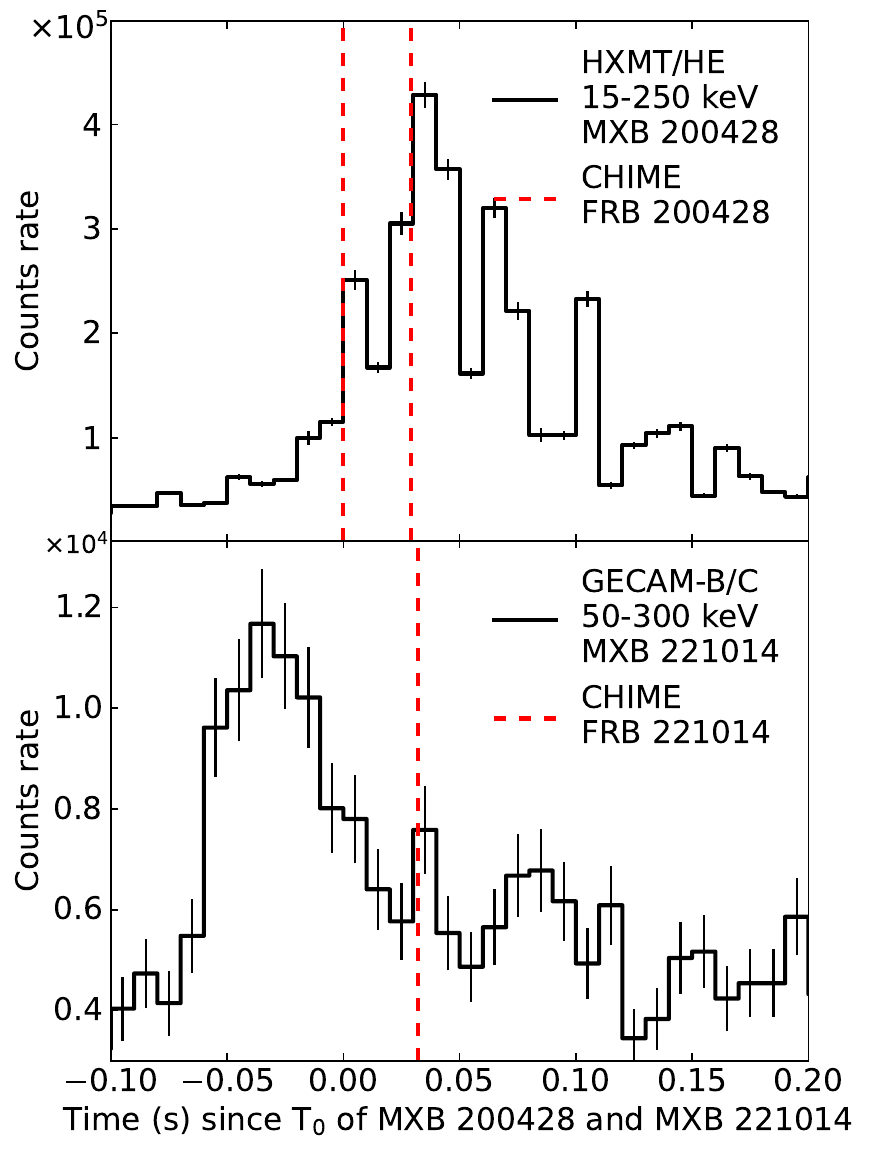}
            \put(-3, 100){\bf b}
        \end{overpic}} \\
        
        \adjustbox{valign=c}{\begin{overpic}[width=0.3\textwidth]{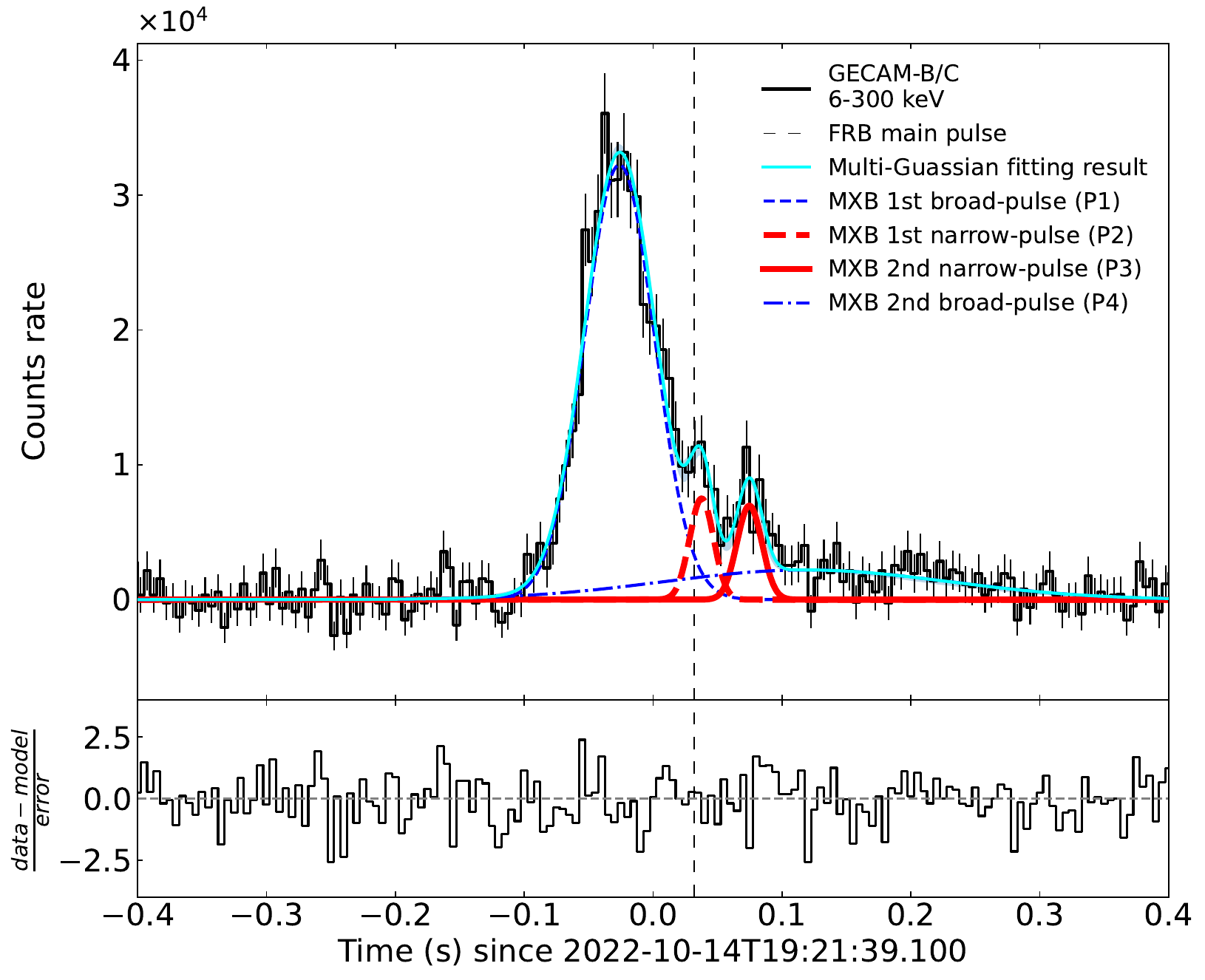}
            \put(-3, 80){\bf c}
        \end{overpic}}
        &
        \adjustbox{valign=c}{\begin{overpic}[width=0.2\textwidth]{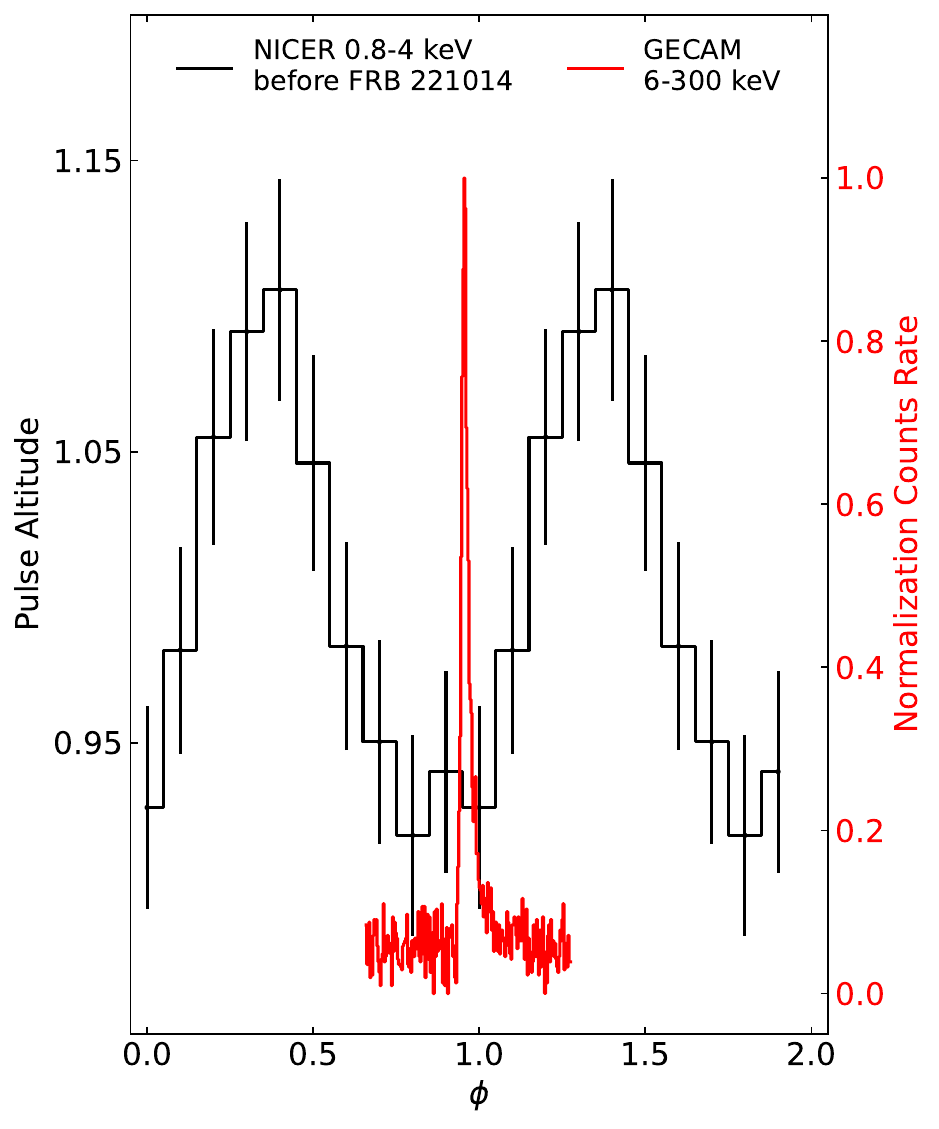}
            \put(-3, 95){\bf d}
        \end{overpic}} \\
    \end{tabular}
\caption{\noindent\textbf{The temporal analysis result of MXB 221014 and FRB 221014.} 
(a), lightcurve of MXB 221014 in different energy range. The dashed line represents the peak time of FRB 221014 \citep{CHIME_221014_arxiv}. A spike coincided with FRB 221014 can be clearly seen at higher energy range. The inset figures are the lightcurve in the unit of counts rate for the same energy range, but with a higher time resolution.
(b), comparison between MXB/FRB 221014 and MXB/FRB 200428. The $T_0$ of MXB/FRB 200428 is 2020-04-28T14:34:24.42650 (geocentric time) while the $T_0$ of MXB/FRB 221014 is 2022-10-14T19:21:39.100. The red dashed lines in top panels are peak time of the two pulse of FRB 200428 \protect\citep{CHIME_200428}. The lightcurve of MXB 200428 is replotted from \protect\citep{HXMT_0428_re}. Both the lightcurve of MXB 200428 and MXB 221014 are not background subtracted in this subfigure. The red dashed lines in bottom panel is the peak time of FRB 221014 \protect\citep{CHIME_221014_arxiv}. Both MXB/FRB 221014 and MXB/FRB 200428 exhibit alignment of narrow pulses, although the narrow x-ray pulses and radio pulses occur during the weak tail region of MXB 221014, while they are around the bright peak region of the MXB 200428. 
(c), fitting of the lightcurve of MXB 221014. The lightcurve of MXB 221014 is combined all the data from GECAM-B and GECAM-C with time aligned. The top panel is the fitting result by four Gaussian function while the bottom panel is the residual. The cyan line is the result of the superposition of the four Gaussian function. Two blue lines are broad pulses with no-radio association and two red lines are narrow pulses. 
The first narrow X-ray pulse (P2) well coincides with the bright main pulse of FRB 221014, while the second narrow X-ray pulse (P3) also aligns with a possible weak radio pulse in CHIME ligutcurve \citep{CHIME_221014_arxiv}. 
(d), the soft X-ray pulse profiles of SGR J1935+2154 obtained with the NICER data before FRB 221014 and the location of MXB 221014 in the soft X-ray pulse profiles, which is roughly aligns with the valley. 
All the lightcurves are aligned in Barycentric Dynamical Time. }
\label{fig:lc_mxb_frb}
\end{figure*}

\begin{figure*}
\centering
\begin{tabular}{ccc}
\begin{overpic}[width=0.25\textwidth]{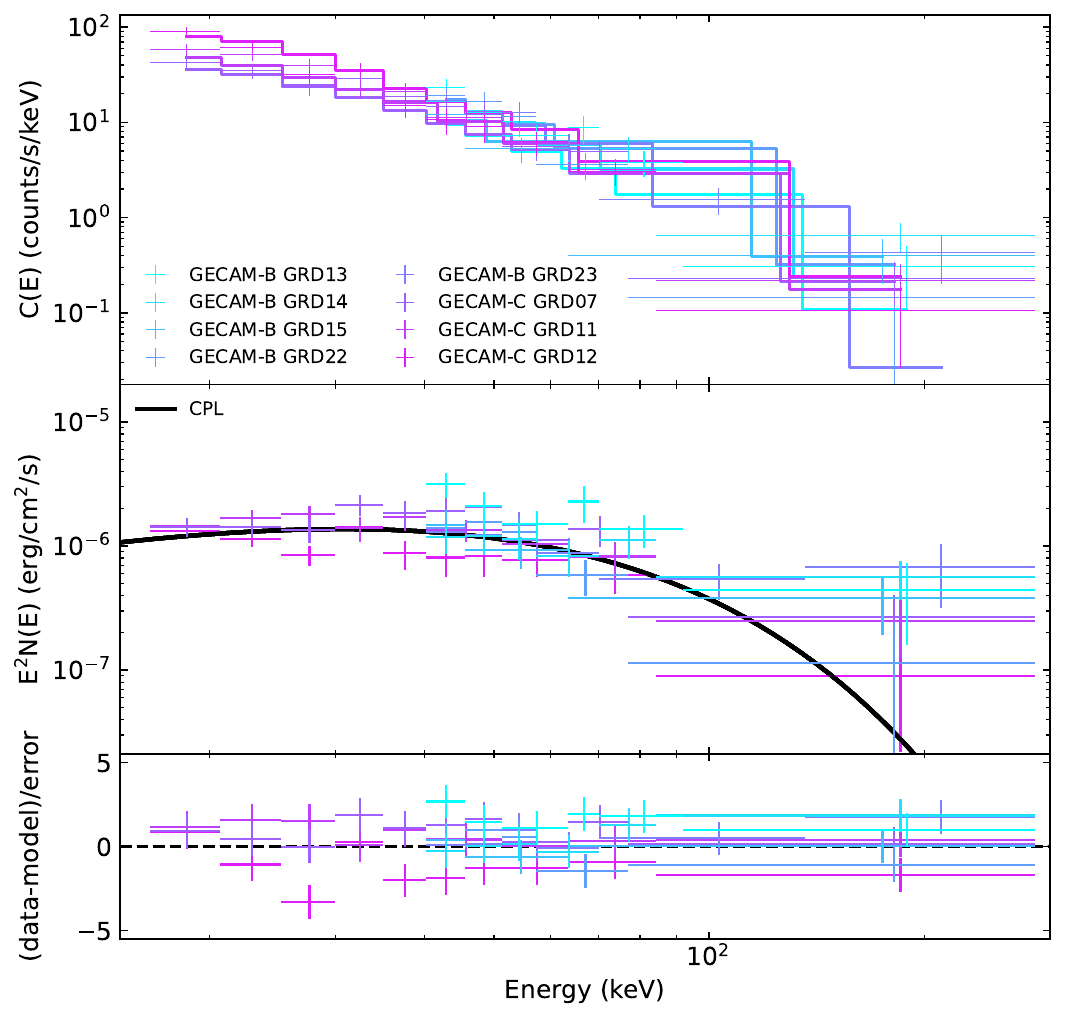}\put(-3, 100){\bf a}\end{overpic} &
        \begin{overpic}[width=0.25\textwidth]{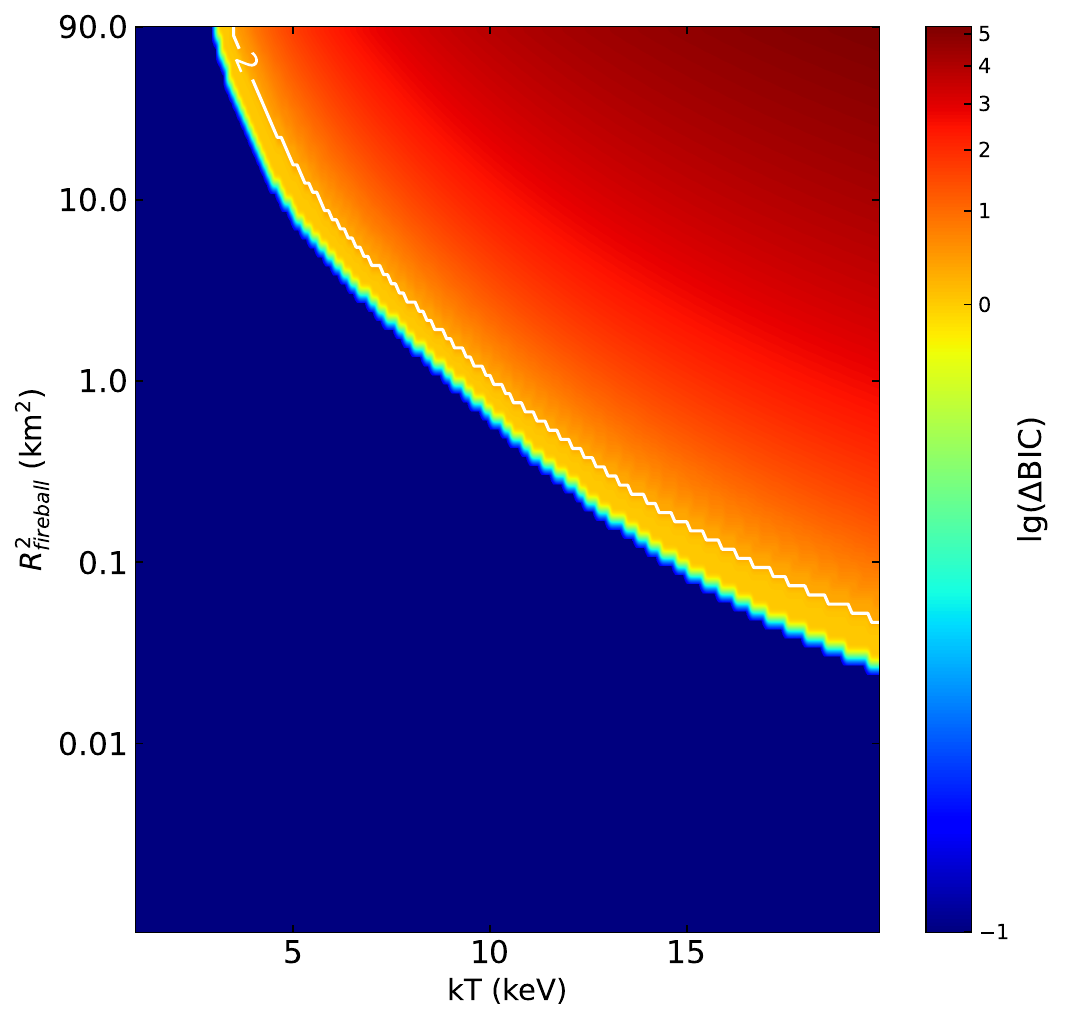}\put(-3, 100){\bf b}\end{overpic}
\end{tabular}
\caption{\noindent\textbf{The spectra and residual of MXB 221014 by utilizing the data from GECAM-B and GECAM-C.} 
(a), the spectra are best fitted with the CPL model in the energy band of 15–300 keV from T$_0$-0.11 s to T$_0$+0.12 s. 
(b), the map of $\Delta$BIC for CPL+BB with different parameters of thermal component. The thermal component with parameters in left bottom region are reasonable for MXB 221014.
}
\label{fig:spec_fitting}
\end{figure*}

\begin{figure*}
\centering
\begin{tabular}{cc}
\begin{overpic}[width=0.5\textwidth]{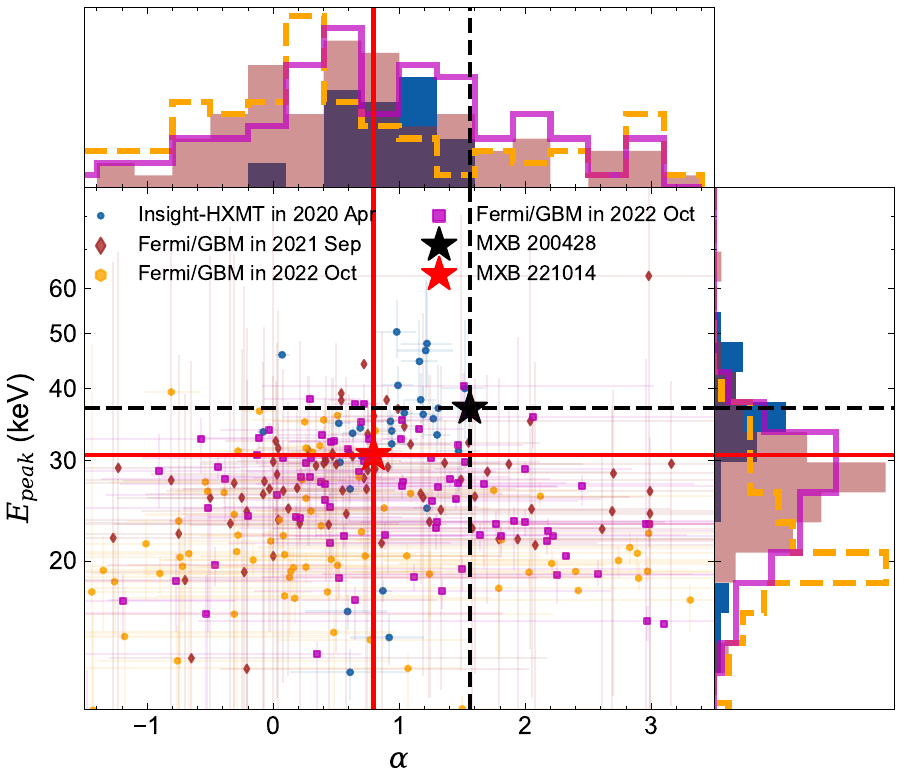}\put(-3, 80){\bf a}\end{overpic} &
        \begin{overpic}[width=0.45\textwidth]{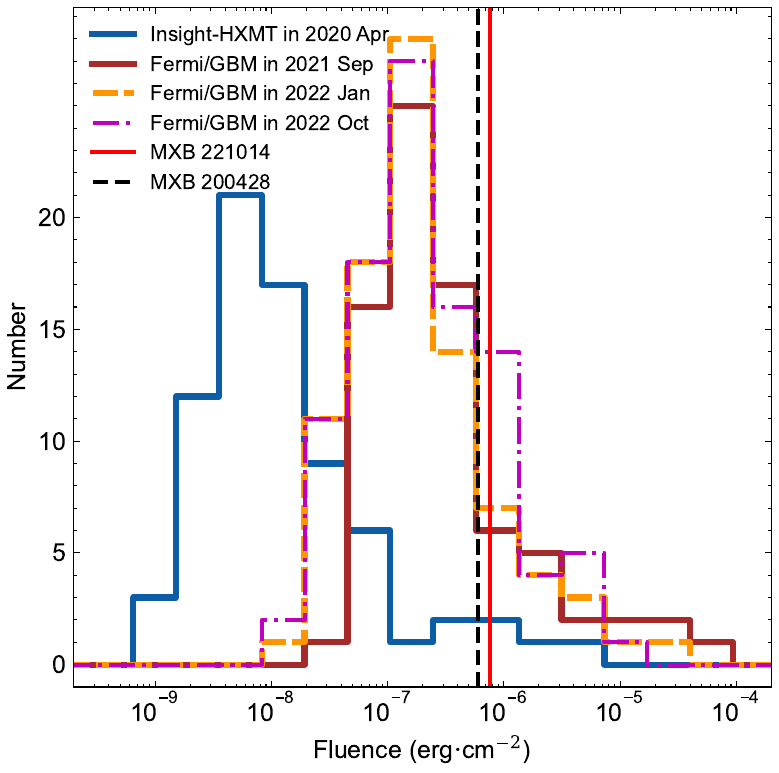}\put(-3, 90){\bf b}\end{overpic}
\end{tabular}
\caption{\noindent\textbf{The comparison of spectral properties of MXB 221014 with other MXB from SGR J1935+2154.} (a), the position of MXB 221014 and MXB 200428 in the $\alpha-E_{peak}$ diagram, indicating MXB 221014 drawn from the same population as the other X-ray bursts from SGR J1935+2154. 
(b), the fluence of MXBs from SGR J1935+2154 compared with burst sample. The position of the two MXBs associated with FRB, ie. MXB 221014 and MXB 200428 are marked as red solid line and black dashed line.}
\label{fig:spec_pars}
\end{figure*}

\begin{figure*}
\centering
\begin{tabular}{cc}
\begin{overpic}[width=0.45\textwidth]{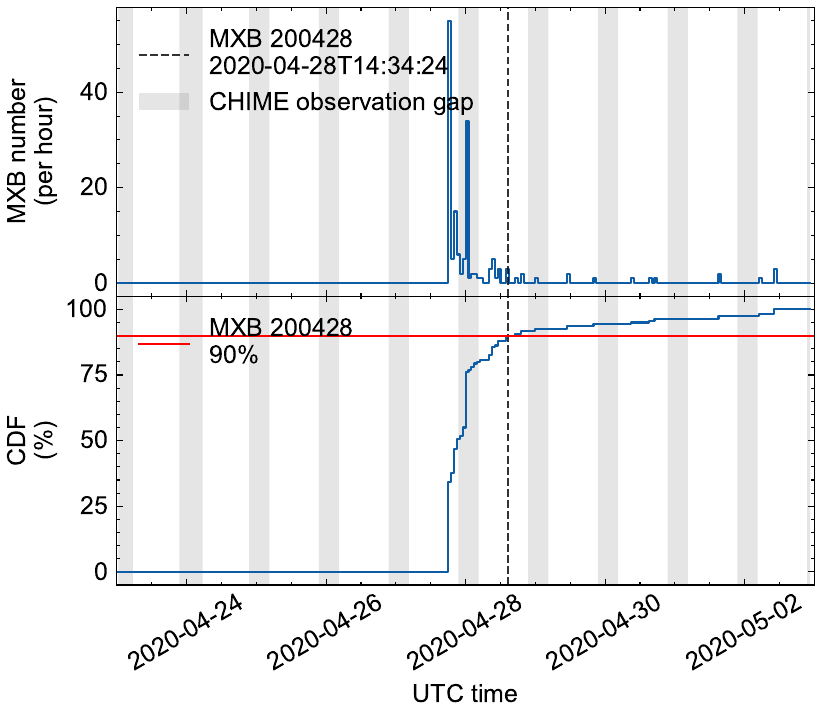}\put(-3, 80){\bf a}\end{overpic} &
        \begin{overpic}[width=0.45\textwidth]{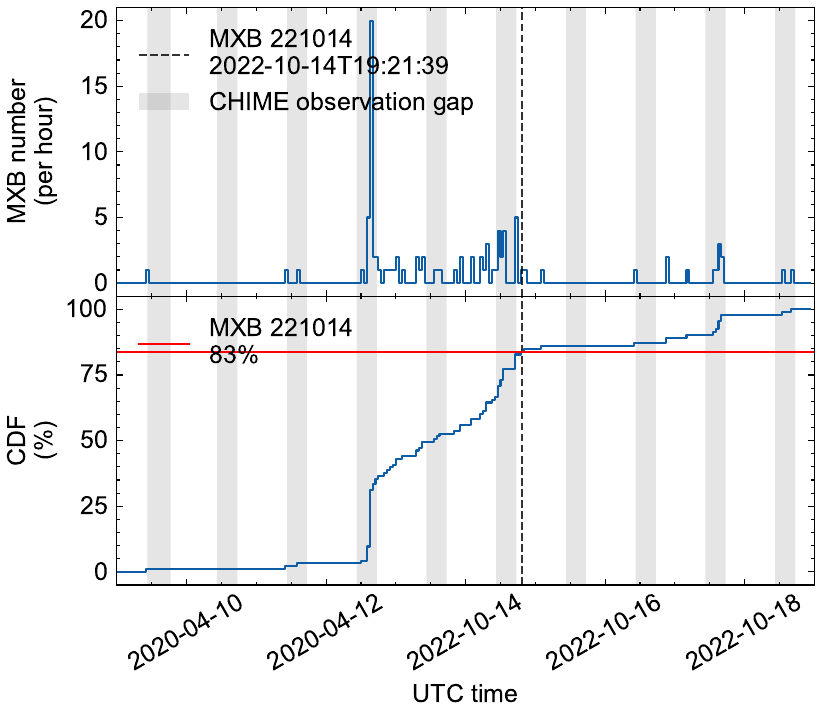}\put(-3, 80){\bf b}\end{overpic}
\end{tabular}
\caption{\noindent\textbf{Distribution of the number of X-ray burst from SGR J1935+J2154 in different activate episode with FRB-associated-MXB.} (a), the activate episode in 2020 April. (b), the activate episode in 2022 October.
The top panels are X-ray burst number detected by GECAM-B, GECAM-C, \textit{Fermi}/GBM and \textit{Insight}-HXMT in each hour. The bottom panels are cumulative distribution. During the time period covered in gray, there was no radio monitoring of SGR J1935+2154 by CHIME due to block of Earth.}
\label{fig:Fig_burst_rate}
\end{figure*}

In this work, we report the GECAM discovery of MXB 221014, the second ever MXB associated with FRB, from the prolific galactic magnetar SGR J1935+2154. 
Thanks to the capability of sensitive monitoring with large field of view, both GECAM-B and GECAM-C triggered and detected MXB 221014, allowing for comprehensive temporal and spectral analysis of this event. 
We find that there is at least a narrow pulse (i.e. P2 in Fig.\ref{fig:lc_mxb_frb}(c)) in MXB 221014 temporally aligned with the radio pulse of FRB 221014, which pins down the association between MXB and FRB, as well as the temporal alignment of narrow X-ray pulse and radio pulse. 
The comprehensive joint analysis of the data from GECAM-B and GECAM-C reveals that a single CPL model is favored for MXB 221014, similar to MXB 200428, although the spectral parameters exhibit relatively ordinary characteristics compared to other X-ray bursts from SGR J1935+2154. 
It is interesting that MXB 221014 has a fluence even higher than MXB 200428, locating in the high-fluence part of the MXB sample. 
But the fireball region emitting thermal spectral component of the burst is also small, resembling the case of MXB 200428.

There are also several differences between MXB/FRB 221014 and MXB/FRB 200428. The narrow pulses of FRB 221014 clearly did not occur at the peak region, but at the tail episode, of MXB 221014, which differs from the case of MXB/FRB 200428. However, it is unclear whether GBT detects any radio pulse around the peak region of MXB 221014. 
Additionally, we find that there is no significant QPO signal in MXB 221014, unlike the case of MXB 200428.

Actually, MXB 221014 is quite ordinary compared to other MXBs samples in terms of spectral and temporal features. 
However, it is still amazing that such a trivial MXB without special properties is associated with a bright radio burst. 
The comparative study of MXB/FRB 200428 and MXB/FRB 221014 provides more constraints for the FRB magnetar origin models developed for the solo event MXB/FRB 200428. 
Besides MXB/FRB 200428 and MXB/FRB 221014, more X-ray bursts associated with radio bursts have been observed (in preparation). 
More questions need to be answered by further observations of more MXB-FRB associated events, e.g. whether the alignment of narrow pulses between MXBs and FRBs is common or not is an interesting question. A continuous observation of SGR J1935+2154 in both X-ray and radio band is necessary, covering both the early and late stages of the active period.

\begin{table*}
\caption{Time alignment of MXB 221014 and FRB 221014}
\begin{tabular}{@{}ccccc@{}}
\toprule
Instrument & Observe Time (ISO) & Observe Time (MJD) & Time delay (ms) & Light travel time (ms)\\
 & UTC & TDB & compared with GECAM-B & compared with GECAM-B\\
\hline
GECAM-B & 2022-10-14T19:21:39.100 & 59866.8081787363 & -- & -- \\
GECAM-C & 2022-10-14T19:21:39.100 & 59866.8081788551 & 10.26 & -10.26 \\
CHIME & 2022-10-14T19:21:39.130 & 59866.8081791061 & 31.95 & -1.95 \\
\hline
\end{tabular}
\label{tab:barycentric_correction}
\\{\raggedright NOTE: (1) This time conversion is achieved with the help of Astropy \citep{astropy:2013, astropy:2018, astropy:2022} and TAT-pulsar \footnote{https://tuoyl.github.io/tat-pulsar/}. 
The J2000 coordinates of the source is set as RA = $\rm 19^h34^m55.5978^s$ and DEC = $\rm +21^\circ53'47.7864''$ \citep{2016MNRAS.457.3448I}. 
And the solar ephemeris for Solar System Barycenter (SSB) correction is DE421, which is the default ephemeride file of TAT-pulsar. \\
(2) The time delay refers to the difference between the observation times of GECAM-C/CHIME and GECAM-B in TDB. 
While the light travel time is the signal delay caused by the distance between different instruments. \\
(3) Observe time of GECAM-B/C is trigger time in UTC, and is converted to TDB directly. 
In fact, MXB 221014 arrive at GECAM-C 10.26 ms earlier than GECAM-B. 
The time delay of 10.26 ms between GECAM-B and C means that GECAM-C triggered 10.26 ms after the trigger of GECAM-B in TDB. 
It is a coincidence that GECAM-B and GECAM-C have the same triggering time, leading to the delay of trigger time in TDB being the same as the light travel time between the two telescope.\\
(4) Observe time of CHIME is arrival time of the peak of main pulse in UTC (topocentric), which is already corrected the effect of dispersion. The topocentric time is converted to geocentric time (59866.8075036782 TT) firstly, then converted to barycentric time (59866.8081791061 TDB).\\}
\end{table*}

\begin{table*}
\caption{\centering{Joint fitting result of $T_0$-0.11\,s to $T_0$+0.12\,s}}
\begin{tabular*}{\hsize}{@{}@{\extracolsep{\fill}}ccccccccc@{}}
\toprule
Model & $\alpha$ & $E_p$ & A$_{\rm NT}$ & kT$_1$ & A$_{\rm TH}$$_1$ & kT$_2$ & A$_{\rm TH}$$_2$ & $\Delta$BIC \\
Unit  &  & keV & photon$\cdot$cm$^{-2}\cdot$s$^{-1}\cdot$keV$^{-2}$ & keV & $R^{2}_{km}$/$D^{2}_{10}$ & keV & $R^{2}_{km}$/$D^{2}_{10}$  & \\
\hline
CPL & 0.80$^{+0.26}_{-0.32}$ & 30.7$^{+2.18}_{-2.68}$ & 46.1$^{+46.50}_{-26.00}$ & - & - & - & - & 0 \\
PL & 2.57$^{+0.05}_{-0.05}$ & - & 4.98$^{+1.05}_{-0.85}\times$10$^3$ & - & - & - & - & 69 \\
BB & - & - & - & 9.30$^{+0.23}_{-0.23}$ & 28.6$^{+3.44}_{-3.01}$ & - & - & 32 \\
BB+BB & - & - & - & 12.2$^{+1.38}_{-1.03}$ & 6.47$^{+4.11}_{-3.02}$ & 4.97$^{+1.07}_{-0.86}$ & 1.71$^{+1.82}_{-0.78}\times$10$^2$ & 6 \\
PL+BB & 2.92$^{+0.40}_{-0.28}$ & - & 7.13$^{+15.7}_{-4.29}\times$10$^3$ & 10.5$^{+0.75}_{-0.70}$ & 10.9$^{+3.68}_{-3.16}$ & - & - & 4 \\
\hline
\end{tabular*}
\label{tab:spec_fitting}
\\{\raggedright NOTE: (1) The PL model is defined $N(E)=AE^{-\alpha}$.\\
(2) The CPL model is defined as $N(E)=AE^{-\alpha}{\rm exp}(-\frac{E}{E_{\rm c}})$. \\
(3) The BB model is expressed as $N(E)=\frac{CAE^2}{{\rm exp}(\frac{E}{kT})-1}$,
where $C=1.0344 \times 10^{-3}$\,photon$\cdot$cm$^{-2}\cdot$s$^{-1}\cdot$keV$^{-2}$, $A=R^2_{km}/D^2_{10}$ is the normalization constant, $R_{km}$ is the source radius in km and $D_{10}$ is the distance to the source in units of 10 kpc, and $kT$ is the temperature in keV. \\
(4) The subscript “NT” means non-thermal, incluing CPL model and PL model while the subscript “TH” means thermal model, which is BB model in this work.\\
}
\end{table*}

\begin{table*}
\caption{\centering{Pulse fitting result of MXB 221014}}
\begin{tabular*}{\hsize}{@{}@{\extracolsep{\fill}}cccc@{}}
\toprule
Pulse ID & Norm (counts rate) & $\mu$ (ms) & $\sigma$ (ms) \\
\hline
P1 & 161.33$^{+3.65}_{-4.13}$ & -26.10$^{+0.73}_{-0.71}$ & 27.04$^{+0.72}_{-0.76}$ \\ 
P2 & 37.67$^{+4.11}_{-4.60}$ & 37.65$^{+1.32}_{-1.49}$ & 9.26$^{+0.55}_{-1.00}$ \\ 
P3 & 35.00$^{+3.48}_{-3.84}$ & 74.75$^{+1.68}_{-1.59}$ & 9.64$^{+0.27}_{-0.59}$ \\ 
P4 & 11.10$^{+1.39}_{-1.37}$ & 117.67$^{+20.00}_{-19.70}$ & 110.56$^{+14.34}_{-13.59}$ \\ 
\hline
\end{tabular*}
\label{tab:pulse_fitting}
\end{table*}

\clearpage

\section*{Acknowledgements}
We appreciate the anonymous reviewer for helpful comments and suggestions. We acknowledge the support by 
the National Key R\&D Program of China (2021YFA0718500), 
the National Natural Science Foundation of China (Grant Nos. 12273042, 
12373047,
12333007,
12303045 
), 
the Strategic Priority Research Program of the Chinese Academy of Sciences (Grant Nos. 
XDA30050000, 
XDB0550300
) and China's Space Origins Exploration Program.
The GECAM (Huairou-1) mission is supported by the Strategic Priority Research Program on Space Science (Grant No. XDA15360000) of the Chinese Academy of Sciences. 
We appreciate the development and operation teams of GECAM and SATech-01 missions. We appreciate Ke-Jia Lee, Heng Xu, Yu-Xiang Huang, Yuan-Pei Yang, Yun-Wei Yu, Zi-Gao Dai, Bing Zhang, and Yuan-Hong Qu for helpful discussions. 
\section*{Data Availability}
The processed data are presented in the tables and figures of the paper, which are available upon reasonable request. The authors point out that some data used in the paper are publicly available at https://gecam.ihep.ac.cn/. 



\bibliographystyle{mnras}
\bibliography{main} 


\bsp	
\label{lastpage}
\end{document}